\documentclass[11pt,a4paper]{article}
\usepackage{geometry}
\usepackage{graphics}
\usepackage{setspace}
\usepackage[scaled=0.9]{helvet}
\usepackage{graphicx}
\usepackage{subfigure}
\usepackage{amssymb}
\usepackage{rotating}
\usepackage{arydshln}
\usepackage{multicol}
\usepackage{abstract}
\usepackage{ucs}
\usepackage[utf8x,utf8]{inputenc}
\usepackage[T1]{fontenc}
\usepackage[swedish, english]{babel}
\usepackage{fancyhdr}

\begin{document}
\bibliographystyle{acm}
\pagestyle{fancy}
\cfoot{\thepage}
\renewcommand{\abstractname}{}

\title{\fontfamily{phv}\selectfont{\huge{\bfseries{A sampling algorithm to estimate the effect of fluctuations in particle physics data}}}}
\author{
{\fontfamily{ptm}\selectfont{\large{Federico Colecchia}}}\thanks{Email: federico.colecchia@brunel.ac.uk}\\
{\fontfamily{ptm}\selectfont{\large{{\it Brunel University, London, Uxbridge, UB8 3PH, United Kingdom}}}}
}
\date{}
\maketitle
\begin{onecolabstract}
Background properties in experimental particle physics are typically estimated using large data sets. However, different events can exhibit different features because of the quantum mechanical nature of the underlying physics processes. While signal and background fractions in a given data set can be evaluated using a maximum likelihood estimator, the shapes of the corresponding distributions are traditionally obtained using high-statistics control samples, which normally neglects the effect of fluctuations. On the other hand, if it was possible to subtract background using templates that take fluctuations into account, this would be expected to improve the resolution of the observables of interest, and to reduce systematics depending on the analysis. This study is an initial step in this direction. We propose a novel algorithm inspired by the Gibbs sampler that makes it possible to estimate the shapes of signal and background probability density functions from a given collection of particles, using control sample templates as initial conditions and refining them to take into account the effect of fluctuations. Results on Monte Carlo data are presented, and the prospects for future development are discussed.
\end{onecolabstract}

\begin{multicols}{2}

\section{Introduction}

While signal and background fractions in a given data set can be obtained using a maximum likelihood estimator, a traditional approach when analysing data in particle physics consists in estimating the shapes of signal and background probability density functions (PDFs) from high-statistics control samples. Generally speaking, such templates can be used to subtract background from a set of candidate events containing signatures of a process of interest. 

However, 
even if the physics processes can be exactly the same, the quantum nature of the underlying physics can produce different features in different events. 
For this reason, the shapes of signal and background PDFs in a collection of events of interest can deviate from the templates obtained using high-statistics control samples. 

When physics analysis leads to the identification of a large enough number of candidate events, such discrepancies are often of no practical relevance. However, in some cases, most notably with reference to searches for new physics, the analysis can result in only a few interesting events, and fluctuations can then 
affect the shapes of signal and background PDFs in ways that cannot be described using high-statistics control samples.


Two major aspects underline the innovative nature of this work. First of all, this article describes the first attempt at estimating the effect of fluctuations on the shapes of probability distributions in particle physics data by using a novel algorithm inspired by the Gibbs sampler \cite{geman}. This builds on the results of previous studies presented in \cite{gibbshep}. The algorithm starts from control sample templates, and uses the stationary distribution of an appropriately-defined Markov Chain to estimate the effect of fluctuations on the shapes of signal and background PDFs. 

Secondly, this paper proposes a new approach to background discrimination. 
Instead of 
concentrating on the likelihood for entire events to contain a 
process of interest, individual particles inside events are assigned a probability for them to originate from signal as opposed to background. This reformulation of background discrimination in terms of a classification problem at the particle level is used 
to illustrate how the algorithm estimates the 
effect of fluctuations on the shapes of particle-level 
PDFs in the data.

However, the two aspects of (i) estimating the effect of fluctuations on PDF shapes and (ii) concentrating on individual particle properties to discriminate signal from background are in principle independent, and can therefore lead to distinct lines of development in the future.

%

For this reason, this research has two main prospective goals. First of all, this work is an initial step toward the development of techniques to study the effect of fluctuations on particle-level probability distributions inside individual events at the Large Hadron Collider (LHC). The emphasis will be on using event-specific templates to associate individual particles with signal or background on a probabilistic basis, taking into account the effect of fluctuations on the shapes of particle-level PDFs event by event.
%
We argue that using different templates for different events to subtract background may improve the resolution of observables of interest and may contribute to reduce systematic uncertainties, depending on the analysis.

Secondly, from a broader perspective, this research aims to guide the future development of new background subtraction techniques based on PDFs that take into account the effect of fluctuations in a collection of events of interest. This aspect is more general and not necessarily restricted to particle-level distributions. Both lines of development refer to the LHC at this point, but we anticipate that they can be of broader relevance.

Finally, it should be noted that it is not the purpose of this article to present the algorithm for use in a proper physics analysis framework at the LHC. On the other hand, this study is a proof of concept of a novel approach to data analysis, illustrating the use of a data-driven technique that estimates the effect of fluctuations on the shapes of probability distributions, with an emphasis on individual particle kinematics.

\section{The algorithm} 

One of the distinctive features of the proposed approach is 
a population-based view of high-energy collision events at the particle level. This aspect 
focuses on 
decomposing a collection of particles 
into subpopulations that are associated with different 
processes and are described in terms of different PDFs. 

The input data set consists of a mixture of particles, some of which originated from a hard scattering of interest, others from background. 
The algorithm uses 
properties of individual particles to decompose the mixture, iteratively sampling from posterior 
distributions that encode information as to which particles are more likely to originate from either process.

As compared to classical mixture models, which typically enforce given subpopulation PDF shapes, our statistical model is based on a nonparametric mixture of the form $F=\sum_{j=1}^K \alpha_j f_j(x)$,
%
%
where the subpopulation fractions $\alpha_j$ (``mixture weights") 
satisfy  
$\sum_{j=1}^K \alpha_j=1$. In the present application, $x$ corresponds to particle pseudorapidity $\eta$, a kinematic variable related to particle polar angle $\theta$ in the laboratory frame in terms of $\eta=-\mbox{ln}(\mbox{tan}\theta/2)$, or to $p_T$ i.e. particle transverse momentum with respect to the beam direction.

The PDFs $f_j$ are here defined in terms of regularized histograms of $x$, namely based on 
spline interpolation of histogram bin contents\footnote{This implementation makes use of the ALGLIB software library \cite{alglib}.}. This aspect is further discussed in \cite{gibbshep2_arxiv} with reference to the existence of a stationary distribution for the associated Markov Chain. 
The symbol $\varphi_j$ 
denotes 
an estimate of PDF $f_j$ corresponding to a splined histogram of $x$ at a given iteration of the sampler. 
Estimates of $f_j$ obtained from control samples 
are
used 
to associate individual particles with signal or background on a probabilistic basis at each iteration of the algorithm (``mapping"). 
The distribution of $x$ in the data set analysed is ultimately estimated 
as a 
splined histogram averaged over iterations. 


In general, given the above mixture $F$ of probability distributions 
and a set of $N$ observations $\{x_i\}_{i=1,...,N}$, 
Markov Chain Monte Carlo (MCMC) techniques 
can be 
used to cluster the latter into $K$ groups by associating each observation with a distribution of origin. 
The Gibbs sampler belongs to this family of numerical methods, and, as anticipated, directly inspired this work.

The pseudocode of the proposed algorithm is shown below, subscripts $sig$ and $bkg$ relating to signal and background, respectively. The value of quantity $v$ at iteration $t$ is denoted by $v^{(t)}$ throughout.

\begin{enumerate}
\item {\bf Initialization:} Set $\alpha_{bkg}=\alpha^{(0)}_{bkg}=\alpha_{sig}=\alpha^{(0)}_{sig}=0.5$, where $\alpha_{bkg}=\alpha_0$ and $\alpha_{sig}=\alpha_1=1-\alpha_{bkg}$. Initial estimates $\varphi^{(0)}_j$ of the subpopulation PDFs $f_j$, $j=1,2$, are given by splined one-dimensional histograms of $\eta$ and $p_T$ obtained from a high-statistics control sample.
\item {\bf Iteration $t$:}
\begin{enumerate}
\item Generate $z_{ij}^{(t)}$ (``allocation variables") 
for all particles $i$ and distributions $j$ 
according to $P(z_{ij}^{(t)}=1 | \alpha_j^{(t-1)}, \varphi_j^{(0)},x_i) = \alpha_j^{(t-1)}\varphi_j^{(0)}(x_i)/\displaystyle\sum_{l=0}^{K-1}  \alpha_{l}^{(t-1)}\varphi_{l}^{(0)}(x_i)$, $K=2$.
Both the nonparametric treatment of the PDFs and the use of $\varphi_j^{(0)}$ instead of $\varphi_j^{(t-1)}$ 
for mapping 
distinguish this implementation from the classical Gibbs sampler for mixture models. 
\item Set $\alpha_j^{(t)}=\sum_i z_{ij}^{(t-1)}/N$, $j=1,2$. 
\end{enumerate}
\end{enumerate}



The sampler 
estimates the shapes of 
the signal and background subpopulations in an input collection of particles, based on the data as well as on initial conditions on the subpopulation PDFs provided by control samples. As opposed to the traditional Gibbs sampler for mixture models, where the stationary distribution of the corresponding Markov Chain relates to the joint posterior of the allocation variables and of the subpopulation PDF parameters, this algorithm 
samples from the posterior probability 
of $z_{ij}$, and PDFs are 
kept fixed at control sample estimates.

This article reports results obtained 
on a Monte Carlo data set generated using Pythia 8.140 \cite{pythia1} \cite{pythia2}. The data set consists of charged particles with $2~\mbox{GeV/c}<p_T<5~\mbox{GeV/c}$ from 50 signal $gg\rightarrow t\bar{t}$ events from pp interactions at $\sqrt{s}=14~\mbox{TeV}$, superimposed on 350 soft QCD interactions, so called Minimum Bias events, to simulate background. 
The data set comprises 1,635 particles, out of which 1,269 originate from signal and 366 from background, corresponding to a fraction of background particles of $\sim0.22$.
In addition to reducing the correlation between particle $\eta$ and $p_T$ to a negligible level, the restriction to the kinematic window $2~\mbox{GeV/c}<p_T<5~\mbox{GeV/c}$ is necessary for $p_T$ to contribute to discrimination between signal and background. If all $p_T$ values were allowed, the difference between signal and background $p_T$ distributions would in fact be too small for $p_T$ to have any discriminating power.
%

The higher-statistics Monte Carlo data set (``control sample") that was used to obtain initial estimates $\varphi_j^{(0)}$ of the subpopulation PDFs $f_j$ consists of a total of $\sim48,000$ particles. It was obtained by generating 1,000 $gg\rightarrow t\bar{t}$ events and superimposing 7 Minimum Bias events per signal interaction. Charged particles were considered in the kinematic range $2~\mbox{GeV/c}<p_T<5~\mbox{GeV/c}$. 
Both samples are exclusively used for illustrative purposes in this article.

The algorithm  
correctly estimated the 
fraction of background particles contained in the input data set 
with a shorter burn-in phase 
than typically reported in the MCMC literature. 
In fact, 
since the sampler essentially refines control sample PDF templates, it already 
starts from reasonable knowledge of the target PDFs. For this reason, the equilibrium distribution of the Markov Chain, although corresponding to an improved description of the PDF shapes that takes fluctuations into account, is normally close to the initial conditions.

The estimated mixture weights can 
be biased 
because 
control sample PDFs are used for mapping instead of the unknown true distributions. 
However, as previously noticed, obtaining the signal and background fractions is not the purpose of this algorithm, and the estimated PDFs were observed to be remarkably unaffected 
by a possible bias on the mixture weights.



\begin{figure*}
\begin{minipage}{25pc}
\centering
\subfigure[]{
\includegraphics[scale=0.35]{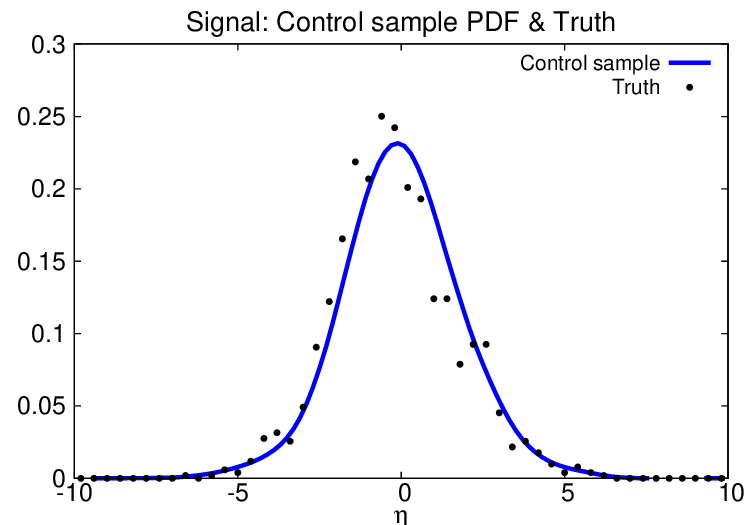}
}
\subfigure[]{
\includegraphics[scale=0.35]{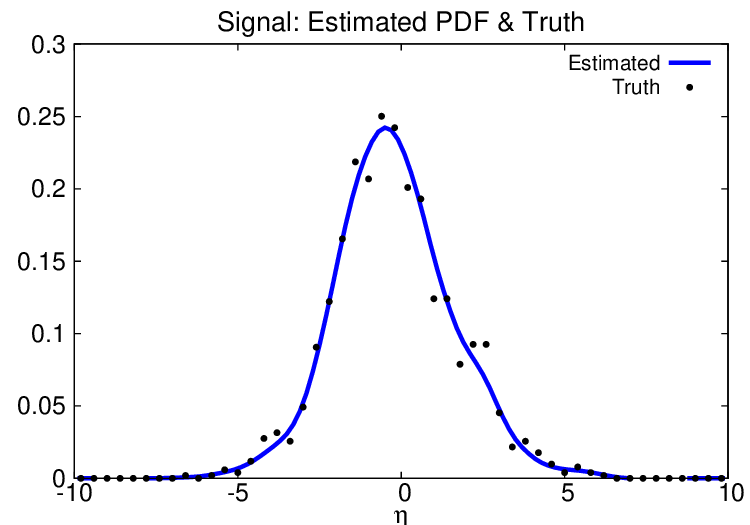}
}\\
\subfigure[]{
\includegraphics[scale=0.35]{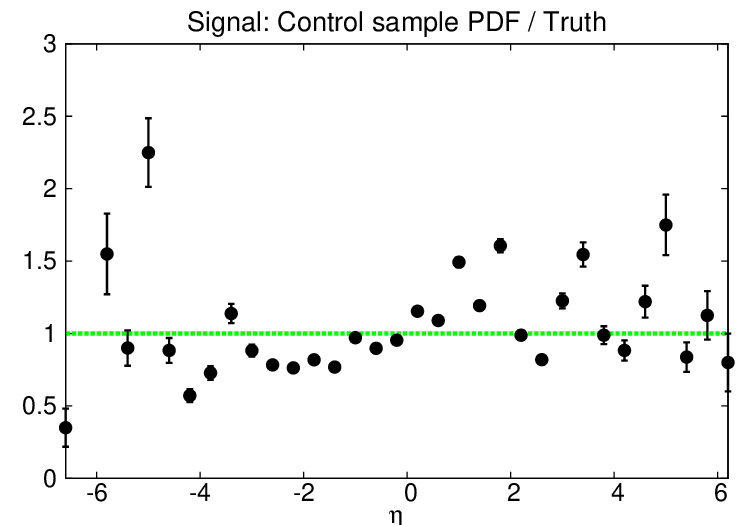}
}
\subfigure[]{
\includegraphics[scale=0.35]{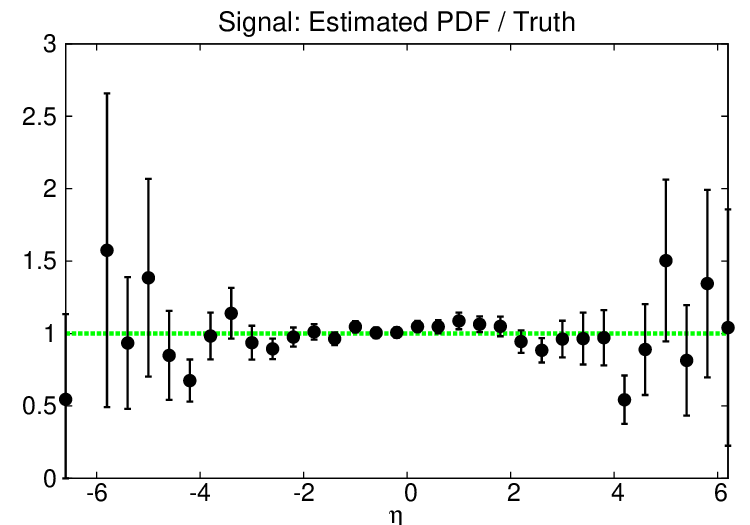}
}
\end{minipage}
\begin{minipage}{12pc}\caption{\label{fig:ds2_eta_sig}(a) True signal $\eta$ distribution normalized to unit area (points) compared to the control sample template (curve). (b) The same true distribution (points) compared to the PDF estimated by the algorithm (curve). (c) Ratio between control sample and true distribution. (d) Ratio between estimated PDF and true distribution.}
\end{minipage}
\end{figure*}

The algorithm was run for 1,000 iterations 
and probabilities were averaged over the last 100. 
No significant gain was found in this study in choosing a higher number of iterations. 

The statistical model is well defined, i.e. the stationary distribution of the Markov Chain exists and is unique, 
which is a required condition for the estimated signal and background PDF shapes to reflect meaningful properties of the data set analysed. This was verified explicitly in two ways. First of all, different initial conditions were used for the mixture weights in the range $[0.1, 0.9]$, and it was confirmed that the estimated signal and background PDFs did not vary appreciably, as expected. Secondly, the algorithm was run repeatedly using different pseudorandom seeds, and the estimated PDFs were again observed to be essentially unaffected.


Figure \ref{fig:ds2_eta_sig} 
illustrates 
the effect of fluctuations in the data by 
comparing the shape of the true signal $\eta$ distribution to the corresponding control sample template (a) and to the PDF estimated by the algorithm (b). The points 
correspond to the true distribution normalised to unit area, while 
the curves result from spline interpolation of 
the 
control sample template (a) and of the 
estimated PDF (b). 
%

The actual distribution in the data 
differs notably from the high-statistics control sample template. Not only is its peak shifted toward lower values, but its shape is not symmetric, as shown by the structure at $\eta\simeq 2$ in figure \ref{fig:ds2_eta_sig} (b). As anticipated, the algorithm can estimate these features thanks to the use of a nonparametric model. 
This underlines the importance of abandoning classical mixture models in favour of a more flexible formulation that can in principle accommodate any deviation in the PDF shapes 
due to fluctuations in the data. 
Figures \ref{fig:ds2_eta_sig} (c) and (d) display the corresponding ratios of control sample to true PDF, and of estimated to true PDF, respectively. 
As it can be noticed, the PDF obtained using the algorithm provides a much more faithful description of the 
true distribution 
than the control sample template. A detailed autocorrelation analysis of the Markov Chain is outside the scope of this work and is left for future studies.
%
Results were also cross-checked on a second Monte Carlo data set corresponding to a different fraction of background particles, as well as on toy Monte Carlo data \cite{gibbshep2_arxiv}.

\section{Conclusions and outlook}
We have described a novel sampling algorithm 
that uses a Markov Chain to estimate the shapes of signal and background PDFs from an input 
data set 
taking fluctuations into account, starting from initial conditions provided by high-statistics control samples. 
This article concentrates on the use of discriminating variables associated with particle-level kinematics. 
It is worth noticing that the algorithm is presented here neither as a classifier nor as a tool for physics analysis at the LHC. This contribution is a proof of concept of a new approach to data analysis in particle physics that aims to estimate the effect of fluctuations on the shapes of probability distributions, while contextually proposing a reformulation of background discrimination as a classification problem at the particle level. 

In particular, this study is a first step toward the development of future mixture decomposition techniques to quantify the effect of fluctuations on the shapes of particle-level probability distributions inside individual events at the LHC. 
We argue that 
using event-by-event templates to subtract background 
may 
improve the resolution of observables of interest, may  
increase the sensitivity in searches for new physics, and may 
reduce systematic uncertainties depending on the analysis. 
Since the number of particles 
in this study 
is 
in line with typical LHC 
charged particle multiplicities, 
these results are an encouraging starting point. 
It should also be noted that, although the present version of the algorithm concentrates on mixtures of one signal and one background subpopulation, this method can in principle be extended to more complex scenarios.

It is worth noticing that efforts to eliminate noise in event-by-event analysis of multiparticle production documented in the heavy ion literature \cite{noise_QGP_01} are fundamentally different from this approach. 
As opposed to traditional techniques whereby functional forms are enforced on the PDFs and parameters are subsequently adjusted within the templates, this approach is based on a nonparametric statistical model that can in principle describe any deviation from the shapes of control sample templates due to fluctuations. Furthermore, this method builds on a 
novel population-based view of particle physics events, 
and reformulates background discrimination as a classification problem at the 
level of individual particles. 


For the sake of completeness, it should be mentioned that a conceptual issue can in principle arise when color connection is involved \cite{gibbshep2_arxiv}, but it is our opinion that this technique can anyway cover many situations of practical relevance. 
It should also be noted that, due to its iterative nature, the algorithm can be outperformed by established methods in terms of execution time. However, the running time 
in this study was $\sim20~\mbox{s}$ on a 2~GHz Intel Processor with 1~GB RAM, and so still reasonable for offline use. 
Improvements may be possible in this respect \cite{gibbshep2_arxiv}. 

\section{Acknowledgments}
The author wishes to thank the High Energy Physics Group at Brunel University for a stimulating environment and for many fruitful discussions. Particular gratitude also goes to the High Energy Physics Group at University College London for their hospitality at an earlier stage, and 
to Prof. Jonathan M. Butterworth for his valuable comments. The author also wishes to thank Prof. Trevor Sweeting and Dr. Alexandros Beskos at the UCL Department of Statistical Science for fruitful discussions. Particular gratitude also goes to 
Prof. Carsten Peterson and to Prof. Leif Lönnblad at the Department of Astronomy and Theoretical Physics at Lund University.


\end{multicols}

\begin{thebibliography}{99}
\bibitem{geman}Geman S and Geman D 1984 {\it IEEE T. Pattern Anal.} {\bf 6} 721
\bibitem{gibbshep}Colecchia F 2012 {\it J. Phys.: Conf. Ser.} {\bf 368} 012031
\bibitem{alglib}ALGLIB (www.alglib.net), Sergey Bochkanov and Vladimir Bystritsky
\bibitem{gibbshep2_arxiv}Colecchia F 2012 ({Preprint} arXiv:1205.5886v1 [physics.data-an])
\bibitem{pythia1}Sjöstrand T, Mrenna S and Skands P 2006 {\it J. High Energy Phys.} JHEP05(2006)026
\bibitem{pythia2}Sjöstrand T, Mrenna S and Skands P 2008 {\it Comput. Phys. Comm.} {\bf 178}
\bibitem{noise_QGP_01}Voloshin S~A, Koch V and Ritter H~G 1999 {\it Phys Rev} C {\bf 60} 024901
\end{thebibliography}
\end{document}